\documentclass[final,numberedheadings,numcites]{aipproc}

\layoutstyle{8x11single}

\usepackage{amssymb}
\usepackage{amsmath}
\usepackage{amsfonts}

\newcommand{\f}{\frac}

\newcommand{\be}{\begin{equation}}
\newcommand{\ee}{\end{equation}}

\newcommand{\D}[2]{\frac{d{#1}}{d{#2}}}

\begin{document}

\title{Neutrinos from charm production in the atmosphere\footnote{Plenary talk given at the Very Large Volume Neutrino Telescope Workshop 2013.}}

\classification{}
\keywords{Atmospheric neutrinos, QCD, cosmic rays}

\author{Rikard Enberg}{address={Department of Physics and Astronomy, Uppsala University, Box 516, S-751 20 Uppsala, Sweden}
}

\begin{abstract}
Atmospheric neutrinos are produced in interactions of cosmic rays with Earth's atmosphere. At very high energy, the contribution from semi-leptonic decays of charmed hadrons, known as the prompt neutrino flux, dominates over the conventional flux from pion and kaon decays. This is due to the very short lifetime of the charmed hadrons, which therefore do not lose energy before they decay. The calculation of this process is difficult because the Bjorken-$x$ at which the parton distribution functions are evaluated is very small. This is a region where QCD is not well understood, and large logarithms must be resummed. Available parton distribution functions are not known at such small $x$ and extrapolations must be made. Theoretically, the fast rise of the structure functions for small $x$ ultimately leads to parton saturation.

This contribution describes the ``ERS'' \cite{ers} calculation of the prompt neutrino flux, which includes parton saturation effects in the QCD production cross section of charm quarks. The ERS flux calculation is used by e.g.\ the IceCube collaboration as a standard benchmark background. We are now updating this calculation to take into account the recent LHC data on the charm cross section, as well as recent theoretical developments in QCD. Some of the issues involved in this calculation are described.
\end{abstract}

\maketitle

%%%%%%%%%%%%%%%%%%%%%%%%%%%%%%%%%%%%%%%%%%%%
%% MAINMATTER
%%%%%%%%%%%%%%%%%%%%%%%%%%%%%%%%%%%%%%%%%%%%

\section{Introduction}

When cosmic rays collide with nuclei in the air in Earth's atmosphere, hadrons are produced in inelastic collisions. These hadrons interact, lose energy, and finally decay, leading to a cascade of particles in the atmosphere that will finally hit the ground. Semileptonic decays of hadrons in the atmosphere generate a flux of neutrinos known as the atmospheric neutrino flux. This is an irreducible background for neutrino observatories such as Super-Kamiokande, IceCube, Antares or the planned KM3NeT. The fluxes at lower energies are pretty well understood, and come mainly from decays of long-lived charged pions and kaons, which are produced in essentially every inelastic collision. This ``conventional'' component of the flux falls steeply with increasing energy, both due to the spectral shape of the incoming cosmic ray flux, and due to the energy loss experienced by the mesons before they decay. Charged pions, for example, have a proper decay length of 8 meters, and with time dilation, their interaction lengths are much smaller than their decay lengths so that they have plenty of time to lose energy. The resulting neutrino energies are therefore downgraded compared to the incoming cosmic ray flux. Theoretical predictions by Honda and collaborators~\cite{honda,gaisserhonda} of the conventional flux agree very well with measurements up to energies of roughly $10^{5}$~GeV~\cite{Schukraft:2013ya}.

At even higher energies, production of hadrons containing charm and bottom quarks leads to another component of the flux known as the ``prompt flux.'' These hadrons are produced much more rarely, but since they decay promptly without losing much energy, the resulting flux is harder and has more or less the same energy dependence as the incoming cosmic ray flux. At energies $\sim 10^{5}-10^{6}$~GeV, this flux is believed to start to dominate the conventional flux, as shown in Fig.~\ref{fig:fluxes}. \footnote{The importance of charm production in atmospheric cascades has been known a long time. We have also recently included charm production in the calculation of neutrino fluxes from astrophysical sources \protect\cite{Enberg:2008jm}.} Coincidentally, this is roughly the same energy that is starting to be probed by the IceCube experiment. The prompt flux is much more poorly known than the conventional flux, due to the theoretical difficulties involved in calculating charm quark production at very high energies. The calculation is sensitive to very small Bjorken-$x$ and very forward rapidities, and this kinematic region is not probed by present collider or fixed target data. One of the most recent predictions of the prompt flux is the ``ERS'' flux of Ref.~\cite{ers}, which was computed before the start of the LHC. This prediction is used by the IceCube experiment as a benchmark for the prompt flux.

The IceCube experiment has recently observed two very high energy neutrino events at about 1~PeV~\cite{Aartsen:2013bka}, and later 26 more events at slightly lower energies~\cite{Aartsen:2013jdh}, all at and above the energies where the prompt flux becomes important. If these events are to be interpreted as (partly) due to a flux of astrophysical neutrinos, it must be clear that they can not be explained by the background coming from atmospheric neutrinos alone. The theoretical uncertainty in the prompt flux would probably not allow assigning the IceCube observations to atmospheric neutrinos---one would need a prompt flux larger than the ERS flux by a factor 15 in order to get a 10\% probability to observe two atmospheric events at 1~PeV~\cite{Aartsen:2013bka}.

However, the significance of their observation depends on the normalization of the atmospheric neutrino flux; the significance of the first two events was $2.8\sigma$, but with an increase in the prompt flux by a factor 3.8, the significance is reduced to $2.3\sigma$. Similarly, the significance of the 28 events to not be completely of atmospheric origin is $4.1\sigma$, which is reduced to $3.6\sigma$ if the prompt flux background is increased by a factor 3.8.
(The factor 3.8 is the level of the measured upper limit on the prompt flux.) Such an increase in the prediction if the various theoretical uncertainties would be better understood is not completely unlikely. On the other hand, if the prediction would become smaller, the significance would increase. In any case, to understand these events it is necessary to understand the background.

The LHC experiments have now measured the charm production cross section at several energies~\cite{LHCcharm}, providing some constraints on the input to the flux calculation. There have also been theoretical developments in QCD after Ref.~\cite{ers}, that can be used to constrain the calculation~\cite{NLL,Albacete:2010sy,Kutak:2012qk}. It is therefore time to improve the calculation to get a better handle on the prediction.

In this contribution I will discuss some of the ingredients in the calculation, and in particular I will discuss why there are large theoretical uncertainties involved. I will necessarily have to skip a lot of details of the calculation, but everything that is not discussed here can be found in our paper \cite{ers} or in the earlier calculations of the prompt flux \cite{Lipari:1993hd,Gondolo:1995fq,Bugaev:1998bi,Pasquali:1998ji,Martin:2003us}. See also the book by Gaisser~\cite{GaisserBook} for details on the description of the cascade in the atmosphere.

\section{Calculation of the neutrino flux}

To find the neutrino flux, we must solve a set of cascade equations that describe the energy loss and decay of nucleons, mesons and leptons in the atmosphere, where the cascade is triggered by an incoming cosmic ray proton. Here I will discuss some of the main points in the flux calculation; for details see Ref.~\cite{ers}.

The flux equations are
\begin{align}
\D{\phi_N}{X} &= -\frac{\phi_N}{\lambda_N} + S( N A \to N Y )
\label{nucleonflux} \\
\D{\phi_M}{X} &= S( N A \to M Y ) -\frac{\phi_M}{\rho d_M(E)}
-\frac{\phi_M}{\lambda_M} + S( M A \to M Y )
\label{mesonflux} \\
\D{\phi_\ell}{X} &= \sum_M S( M \to \ell Y )
\label{leptonflux}
\end{align}
where $\ell = \mu,\nu_\mu,\nu_e$ and the mesons include unstable baryons:\ for
prompt fluxes from charm $M= D^\pm$, $D^0$, $\bar D^0$, $D_s^\pm$,
$\Lambda_c^\pm$. $d_M=c\beta\gamma\tau$ is the decay
length and $\lambda_{i}$ are the interaction lengths for hadronic energy loss. The variable $X$ is the slant depth, essentially the amount of atmosphere that a given particle has traversed.
The initial conditions for the fluxes are zero for all but the nucleon flux, which is given by the incoming cosmic ray flux, i.e., we assume the cosmic ray flux to be composed of protons. We use a primary nucleon flux parametrization with a knee from Ref.~\cite{Gondolo:1995fq}.

The functions $S( k \to j )$ are the regeneration functions given by
\be
S( k \to j ) = \int_E^\infty dE' \f{\phi_k(E')}{\lambda_k(E')}
\D{n(k \to j ;E',E)}{E},
\ee
where $E'$ and $E$ are the energies of the incoming and outgoing particle. This is where the production cross sections and decay matrix elements come in. To solve the flux equations, we use the semi-analytic method of $Z$-moments used e.g.\ in\cite{GaisserBook,Lipari:1993hd,Gondolo:1995fq,Pasquali:1998ji}. This is known to be a good approximation (see e.g.\ \cite{Gondolo:1995fq}). The $Z$-moments are defined by
\be
Z_{kh}=\int_E^\infty dE'
\frac{\phi_k(E',X,\theta)}{\phi_k(E,X,\theta)}\frac{\lambda_k(E)}{\lambda_k(E')}
\D{n(k A \to h Y;E',E)}{E},
\ee
which after assuming that the energy- and $(X,\theta)$-dependence of the flux is factorized as
$\phi_k(E,X,\theta)=E^{-\gamma-1}\phi_k(X,\theta)$ takes the form
\be
Z_{kh}=\int_E^\infty dE'
\left(\frac{E'}{E}\right)^{-\gamma-1} \frac{\lambda_k(E)}{\lambda_k(E')}
\D{n(k A \to h Y;E',E)}{E}, \label{Zkh}
\ee
so that only the energy dependence of the flux enters the $Z$-moments.

The $Z$-moments $Z_{NN}$ and $Z_{MM}$ as well as the interaction lengths $\lambda_{i}$ describe energy loss and scattering in the atmosphere and are computed using parametrized scattering cross sections (see \cite{ers} for details). The charm production process is described entirely by the $Z$-moment $Z_{NM}$, for all charmed mesons $M$. The decay $Z$-moments $Z_{M\ell}$, finally, are computed using two- or three-body phase space according to Refs.~\cite{Lipari:1993hd,Bugaev:1998bi}. Note that all $Z$-moments depend on the energy.

The cascade equation for the mesons can then be rewritten in terms of the $Z$-moments as
\be
\D{\phi_M}{X} = - \frac{\phi_M}{\rho d_M} - \frac{\phi_M}{\lambda_M}
+Z_{MM} \frac{\phi_M}{\lambda_M} + Z_{NM}  \frac{\phi_N}{\lambda_N}
\label{phiM}
\ee
with simpler equations for the nucleon and lepton fluxes. Eq.~(\ref{phiM}) is solved by obtaining separate solutions in the high- and low-energy limits where the interaction or decay terms dominate, respectively.
The full solution is then obtained as an interpolation between the high- and low-energy solutions.

The two solutions are separated by a critical energy $\epsilon_{M}$, which is different for different mesons, and that additionally depends on the zenith angle, since the amount of atmosphere the cascade traverses depends on this angle. The equations for the lepton fluxes are
\begin{align}
\phi^\text{low}_\ell &= Z_{M\ell,\gamma+1} \frac{Z_{NM}}{1-Z_{NN}}\phi_N(E) \\
\phi^\text{high}_\ell &= Z_{M\ell,\gamma+2} \frac{Z_{NM}}{1-Z_{NN}}
\frac{\ln(\Lambda_M/\Lambda_N)}{1-\Lambda_N/\Lambda_M} \frac{\epsilon_M}{E}
\phi_N(E),
\end{align}
where $(\gamma+1)$ is the spectral index of the incoming cosmic ray flux at high and low energy and $Z_{M\ell,\gamma+1}$ and $Z_{M\ell,\gamma+2}$ are calculated using these fluxes. The attenuation lengths $\Lambda_i$ are defined as
$\Lambda_N(E)={\lambda_N(E)}/{(1-Z_{NN}(E))}$, etc. The lepton fluxes are thus proportional to the cosmic ray flux, but the energy dependence is modified by the energy dependence of the $Z$-moments and the attenuation lengths, and in addition the high-energy flux is suppressed by one power of energy compared to the cosmic ray flux. This additional factor of the energy comes from the gamma factor in the decay length, and the suppression only becomes effective when the charmed hadrons start losing energy before they decay, at very high energy.

\section{Charm production in QCD}
The $Z_{NM}$ functions can be rewritten in terms of the energy fraction $x_{E}=E/E'$ of the charmed particle. The essential ingredient is the differential cross section for production of a charm quark pair ${d\sigma(pp \to c\bar c)}/{dx_E}$. In the calculation of the neutrino flux we actually compute the cross section for production of charmed hadrons $M$ by convoluting the charm quark cross section with the relevant fragmentation functions, but let us first for simplicity consider the charm quark cross section by itself. Then $x_E=E_c/E_p$. At high energy, we have $x_E\simeq x_F$, where $x_{F}$ is the more convenient Feynman variable. In perturbative QCD, the dominant contribution to the cross section comes from the subprocess $gg\rightarrow c\bar{c}$, and the cross section is then given by
\begin{equation}
\frac{d\sigma}{dx_F}=\int
\frac{ dM_{c\bar{c}}^2}{(x_1+x_2) s}
\sigma_{gg\rightarrow c\bar{c}}(\hat{s}) G(x_1,\mu^2) G(x_2,\mu^2)
\end{equation}
where $x_{1,2}$ are the momentum fractions of the gluons, $x_F=x_1-x_2$ is the Feynman variable, and
$G(x,\mu^2)$ is the gluon distribution of the proton, The center-of-mass energy of the partonic system is given by $\hat s=x_{1} x_{2} s$, and $\mu$ is the factorization scale.
Given the charm--anticharm invariant mass $M_{c\bar{c}}$, the
fractional momenta of the gluons, $x_{1,2}$, can be expressed in terms of the
the Feynman variable, $x_F$,
\be
\label{eq:x12}
x_{1,2} = \frac{1}{2}\left( \sqrt{x_F^2+\frac{4M_{c\bar c}^2}{s}}
\pm x_F\right) \ .
\ee
Typically the factorization scale is taken to be of the order of $2m_c$.

The squared center-of-mass energy is $s=2E_{p}m_{p}$, so from
Eq.\ (\ref{eq:x12}), it is clear that at high energy the dominant contribution is the highly asymmetric case where one gluon PDF is evaluated at $x_1 \sim x_F$ and the other at $x_2\ll 1$.

To illustrate the $x$-values involved, for an incoming energy of $E_{p}=100$~TeV and assuming the charm quarks do not have appreciable relative $p_{T}$, we get for
$x_{F}=0$ (central production) that $x_2 = 5\times 10^{-3}$, and for $x_F=1$ (forward limit) we get $x_2 = 3\times 10^{-5}$. For $E_{p}=1$~PeV, we instead get $x_2 = 2\times 10^{-3}$ and $x_2 = 3\times 10^{-6}$, respectively. These are extremely small values of $x$---for forward production they are far smaller than anything accessible at today's accelerators.

The gluon distribution cannot be measured directly and has large uncertainties at small $x$, and especially so for the low factorization scales $\mu\sim 2 m_c$ that we are interested in. The evolution of the PDFs at small $x$ involves large logarithms $\alpha_{s}\log(1/x)$ that need to be resummed. This is done by solving the BFKL equation \cite{bfkl}, which predicts a rapid power growth of the PDF as $x\to 0$. However, there are no measurements at very small $x$ values, so the behavior of the PDFs is not well known. The commonly used PDF parametrizations have quite small minimum $x$ values, but the shape of the PDF there is an extrapolation from data at higher $x$. For example, the PDF fits from the HERA experiments have $x > 10^{-4}$. The LHC experiments will reach smaller $x$, but at much larger factorization scales. (The ideal machine to measure PDFs at very small $x$ would probably be the proposed LHeC electron--proton collider~\cite{AbelleiraFernandez:2012cc}.)

\begin{figure}[t]
\includegraphics[width=0.48\columnwidth]{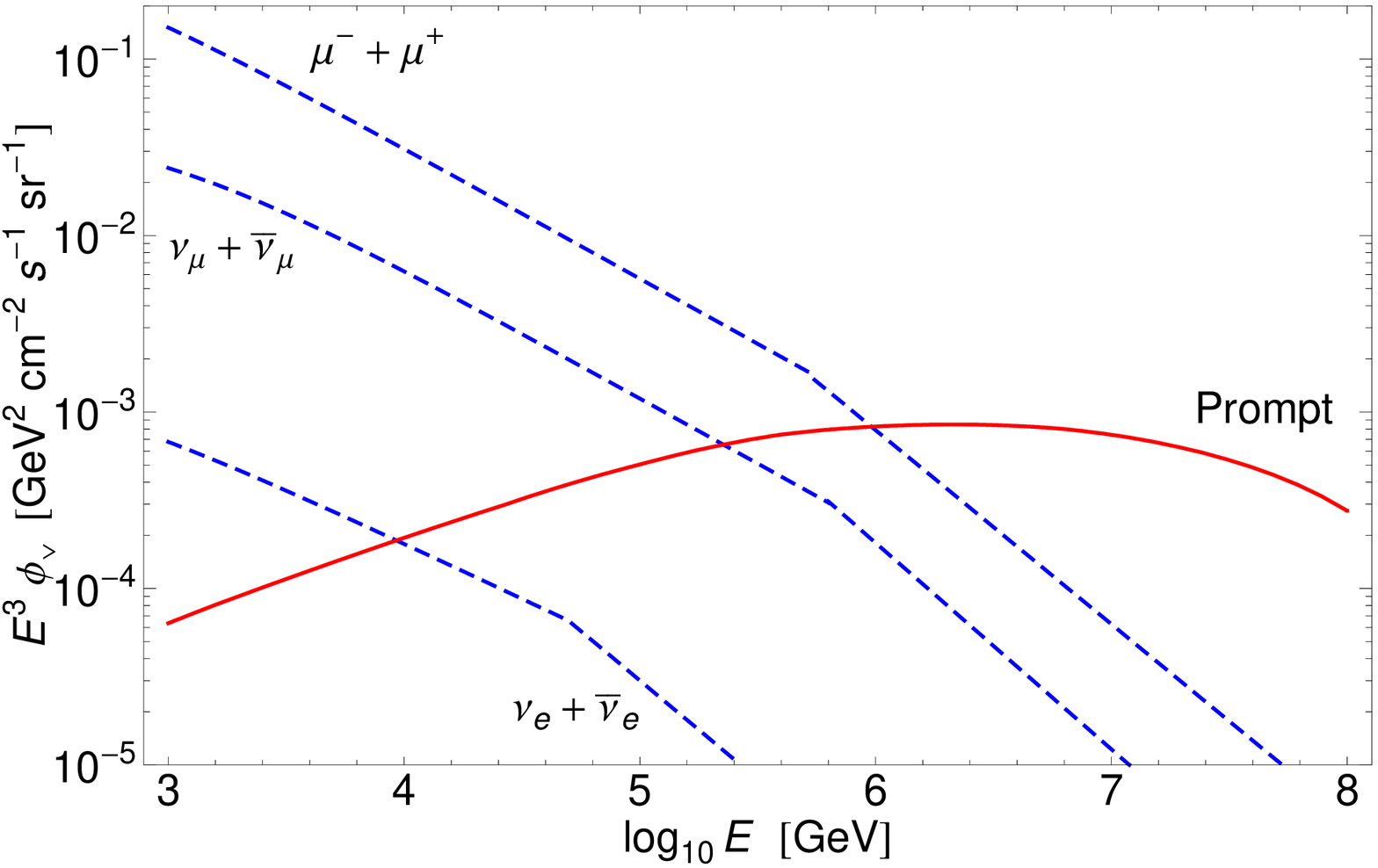} \quad
\includegraphics[width=0.48\columnwidth]{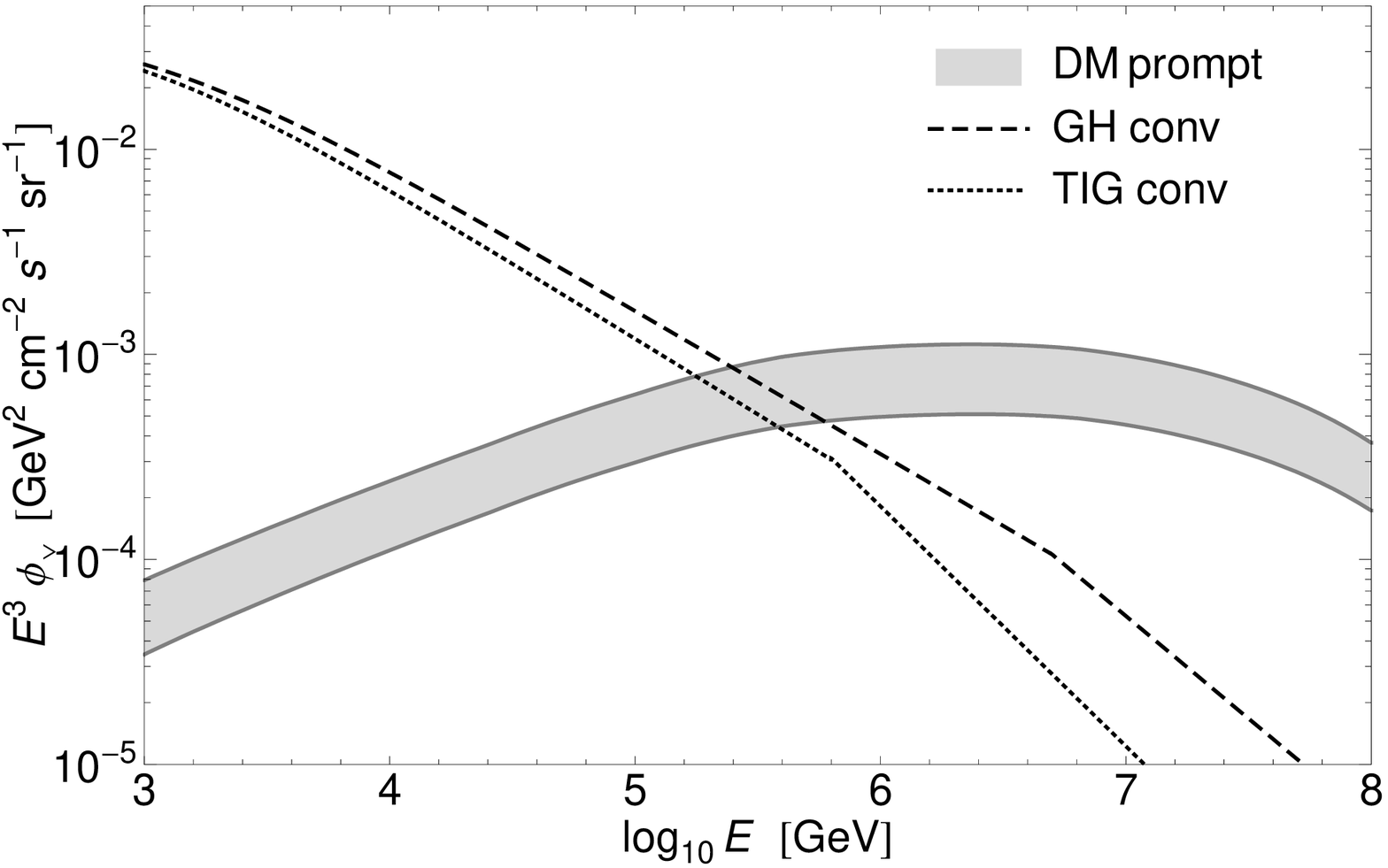}
\caption{Left: Prompt and conventional fluxes
of $\nu_\mu + \bar\nu_\mu$, $\nu_e + \bar\nu_e$, and $\mu^+ + \mu^-$ in the vertical direction. Conventional fluxes from Thunman, Ingelman and Gondolo (TIG), Ref.~\protect\cite{Gondolo:1995fq}.
The three prompt
fluxes are approximately equal, so only the
$\nu_\mu + \bar\nu_\mu$ flux is shown.
Right: Prompt and conventional $\nu_\mu + \bar\nu_\mu$ fluxes in the vertical
direction. The shaded band
is the theoretical uncertainty band for the prompt flux from \cite{ers}.
Conventional fluxes from Gaisser and Honda (GH)
\cite{gaisserhonda} and from TIG. Figures from Ref.\ \cite{ers}.
\label{fig:fluxes}}
\end{figure}

The rapid growth of the gluon PDF at small $x$ can be interpreted as a growth in the number density of gluons. When the density becomes large enough, unitarity can be violated, but taking into account that gluons may begin to recombine at large densities, unitarity is saved. This leads to a reduction in the growth at small $x$. This phenomenon is known as parton saturation, and would reduce the gluon density and thus the cross section. The full description of parton saturation is complicated~\cite{MV,JIMWLK,Balitsky,Gelis:2008rw,Kovchegov}, but there is a sort of ``mean-field approximation'' to the full description known as the Balitsky--Kovchegov (BK) equation~\cite{Balitsky,Kovchegov}, which is phenomenologically very useful. In \cite{ers} we used an approximate solution of the BK equation due to Iancu, Itakura and Munier~\cite{Iancu:2003ge}, which has a handful of free parameters that have been fitted to HERA data~\cite{Iancu:2003ge,Soyez:2007kg}. We performed the calculation in a framework known as the dipole picture of small-$x$ QCD. Due to space limitations I will not describe this theoretical framework here, but refer to \cite{ers} and references therein. The main point is that we are taking the effects of parton saturation into account, which leads to a reduction of the cross section compared to a fixed order QCD calculation. The calculation is done using a different way of factorizing the cross section, with the essential ingredient being the dipole cross section $\sigma_\text{dip}$ that describes the scattering of a quark--antiquark pair on the nucleon or nucleus.

However, as there is limited knowledge experimentally of the behavior of the PDFs at small $x$, it is not known how saturation works and manifests itself---or indeed if it occurs at all. There is thus a substantial theoretical uncertainty in the calculation of the charm cross section. The formalism we are using has been tested against DIS data from HERA at small $x$~\cite{Iancu:2003ge,Soyez:2007kg}. In Ref.~\cite{Goncalves:2006ch}, charm production in hadron--hadron collisions was calculated in the same framework we are using, and was tested against the limited amount of available data on this process. But we need much smaller $x$, so the agreement at larger $x$ does not necessarily mean that the extrapolation works well for smaller $x$.

There are two ways of improving the result: there is now data on charm production from the LHC available~\cite{LHCcharm}, and there have been some theoretical developments in improving the predictions in small-$x$ QCD. It is well-known that the BFKL equation that describes the small-$x$ evolution without saturation must be supplemented by next-to-leading logarithmic corrections to give a stable and phenomenologically sound result. This should also be incorporated in the BK equation, which is essentially the BFKL equation minus a non-linear term, but this must be done approximately, perhaps along the lines of \cite{NLL,Albacete:2010sy,Kutak:2012qk,Cazaroto:2011qq}.

There are also inherent uncertainties in the dipole model saturation calculation related to parameter values, choices of parametrization of the gluon distribution, of the factorization scale, and of the treatment of quark fragmentation into hadrons. We have quantified these by varying them within reasonable limits. In particular we vary the factorization scale between $\mu_F=2 m_c$ or $\mu_F=m_c$, and the charm quark mass between $m_c=1.3$~GeV and $m_c=1.5$~GeV. We also choose a few different available gluon PDFs and two different quark fragmentation functions. This gives rise to the uncertainty band shown in the right plot in Fig.~\ref{fig:fluxes}. The shape of the  neutrino flux does not depend strongly on these choices, but the overall normalization varies by up to a factor of two.

\begin{figure}[t]
\includegraphics[width=0.6\columnwidth]{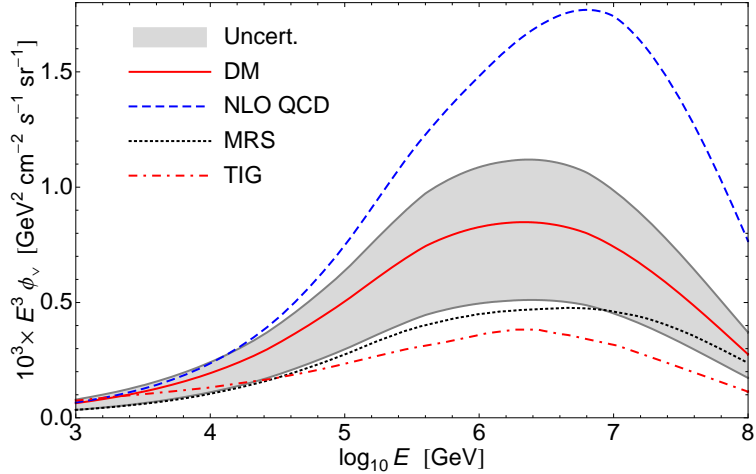}
\caption{Prompt muon neutrino fluxes.
The shaded area is the theoretical uncertainty of the ERS result, as discussed in the text.
The solid line in the band is our standard result.
The dashed curve is the NLO QCD calculation of
Ref.~\protect\cite{Pasquali:1998ji} (PRS), modified to include fragmentation functions
The dotted  curve is the saturation model result of Ref.~\protect\cite{Martin:2003us} (MRS).
The dash-dotted curve is
the LO QCD calculation of Ref.\ \protect\cite{Gondolo:1995fq} (TIG). Figure from Ref.~\cite{ers}.
\label{fig:compare_others}}
\end{figure}

In Ref.~\cite{ers}, we also compared our flux prediction to several earlier evaluations of the flux, see Fig.~\ref{fig:compare_others}. There is a range of predictions with a spread of about a factor of six. The NLO QCD calculation of Ref.~\cite{Pasquali:1998ji} is a regular QCD calculation that does not include saturation but uses a power-law extrapolation of the gluon PDF to small $x$. It is larger than our prediction by roughly a factor two. The toy model saturation prediction of  Ref.~\cite{Martin:2003us} is smaller by a similar factor.

It is clear that if saturation occurs, the cross section is smaller than it would be if saturation does not occur. The NLO QCD calculation can therefore be seen as an upper limit on the neutrino flux.\footnote{Note that if saturation does occur at the scales probed in existing experiments, then it is effectively included in the existing PDF fits, which makes it hard to discern from other effects.} If saturation does occur, as is expected on theoretical grounds, the ERS result is not likely to be much smaller than a new calculation, but it is not known what result an improved calculation might give. It is known, however, that when next-to-leading corrections to saturation are included, the growth of the cross section with energy is further suppressed~\cite{NLL}.

To obtain some more information on these issues, we are planning to update the ERS prediction both with an improved saturation calculation that includes next-to-leading logarithmic corrections, and with a NLO QCD calculation with more modern PDF fits.

%%%%%%%%%%%%%%%%%%%%%%%%%%%%%%%%%%%%%%%%%%%%%%%%
%% BACKMATTER
%%%%%%%%%%%%%%%%%%%%%%%%%%%%%%%%%%%%%%%%%%%%%%%%

\begin{theacknowledgments}
Everything described in this talk was done in collaboration with my friends Mary Hall Reno and Ina Sarcevic, and I thank them for a very fruitful collaboration. I also thank the organizers of the VLVnT 2013 workshop for the invitation to present this talk.
\end{theacknowledgments}

\end{document}